\documentclass[%
reprint,
nobibnotes,
 amsmath,amssymb,
 showkeys,
 aps,
pre,
]{revtex4-1}
\usepackage{graphicx}
\usepackage{dcolumn}
\usepackage{bm}
\usepackage{hyperref}
\usepackage{bigints}
\usepackage{booktabs}
\usepackage{color}
\usepackage{ragged2e}
\usepackage{epstopdf}
\usepackage{multirow}

\begin{document}
\title{Two-dimensional time-reversible ergodic maps with provisions for dissipation}
\author{Puneet Kumar Patra}
\affiliation{Advanced Technology Development Center, Indian Institute of Technology Kharagpur, West Bengal, India 721302}
\email{puneetpatra@atdc.iitkgp.ernet.in}

\date{\today}

\begin{abstract}
A new discrete time-reversible map of a unit square onto itself is proposed. The map comprises of piecewise linear two-dimensional operations, and is able to represent the macroscopic features of both equilibrium and nonequilibrium dynamical systems. Our operations are analogous to sinusoidally driven shear in the two dimensions, and a radial compression/expansion of a point lying outside/inside a circle centered around origin. Depending upon the radius, the map transitions from being ergodic and nondissipative (like in equilibrium situations) to a limit cycle through intermediate multifractal situations (like in nonequilibrium situations). All dissipative cases of the proposed map suggest that the Kaplan -- Yorke dimension is smaller than the embedding dimension, a feature typically arising in nonequilibrium steady-states. The proposed map differs from the existing maps like the Baker's map and Arnold's cat map in the sense that (i) it is reversible, and (ii) it generates an intricate multifractal phase-space portrait. 
\end{abstract}

\keywords{Ergodic maps; Dissipative maps, Random number generator, Lyapunov exponents}

\maketitle

\section{Introduction}
{A large class of nonequilibrium systems can be modelled using time-reversible dynamics that allow thermodynamic and hydrodynamic dissipation \cite{hoover_book_12}. Such models usually augment} the Newtonian (or equivalently the Hamiltonian) equations of motion with friction like terms in a manner that energy can be exchanged with the surroundings \cite{nose_84,hoover_85,hh_thermostat,tuckerman_92,pb_thermostat, c12_thermostat}. If $q$, $p$ and $\xi$ denote the position, momentum and friction-like variables, then the time reversibility of the underlying equations of motion tell us that both the forward trajectory $(q,p,\xi)$ and the time-reversed trajectory $(q,-p,-\xi)$ satisfy the evolution equations \cite{hoover_96}. The dynamical systems must be ergodic i.e. a trajectory initiating at a particular microstate must visit, given sufficient time, the neighbourhood of all allowable microstates \cite{patra_nonergodicity14,patra_hoover_15}. Ergodicity of the dynamics (i) makes the time-averages of a dynamical property independent of the initial conditions, and (ii) enables us to equate the phase-space averages with the time averages, rendering the applicability of the molecular dynamics simulations. 

It has been known for a while now that the equilibrium systems are characterized by conservative dynamics and the nonequilibrium dynamical systems are invariably characterized by dissipative dynamics caused by the conversion of useful work and internal energy into heat, as mandated by the second law of thermodynamics \cite{evans_book, hoover_97}. Additionally, the dissipative continuous dynamical systems have intricate multifractal dynamics, with their Kaplan-Yorke dimension \cite{kaplan-yorke} smaller than the embedding dimension, so that the entropy becomes singular and divergent \cite{hoover_13}. Thus, the dynamics of a nonequilibrium system follows a unidirectional ``arrow of time'', something seemingly in contradiction with the time-reversible nature of the dynamics. The paradox of time-reversibility of the dynamics and the arrow of time may be resolved by arguing the stability of the multifractals obtained in the forward trajectory relative to the time-reversed trajectory -- the forward trajectories are Lyapunov stable i.e. $\sum L_i < 0$, while the reverse trajectories are Lyapunov unstable i.e. $\sum L_i > 0$. Here, $L_i$ represents the $i^{\text{th}}$ Lyapunov exponent. Thus, the strange attractor of the forward trajectory acts as a sink whose source is provided by the repeller of the time-reversed trajectory. 

However, analyzing and understanding the physics of nonequilibrium dynamics even for simple three-dimensional and four-dimensional cases is not trivial. For example, ``holes'' embedded within a subspace of a four dimensional system \cite{patra_nonergodicity14}, or near zero-measure tori in three dimensional systems \cite{hoover-sprott-patra-15} are very challenging to determine. A simple alternative to understanding the behavior of these complex dynamical systems is to study the equivalent two-dimensional maps. 

Ideally, these maps should show all the features of a continuous dynamical system discussed before: (i) time reversible, (ii) ergodic, and (iii) allow dissipation to occur. We define time-reversibility of the maps analogous to the time reversibility of the continuous dynamical systems \cite{hoover_96}: consider a mapping operation, $\mathcal{M}$, that maps a point $(q,p)$ to another point $(q^\prime, p^\prime)$, upon reversing $p^\prime \to -p^\prime$ and performing another mapping operation results in the original state but with reversed $p$: $(q^\prime,-p^\prime) \xrightarrow{\mathcal{M}} (q,-p)$. 

\begin{figure}
\includegraphics[scale=0.35]{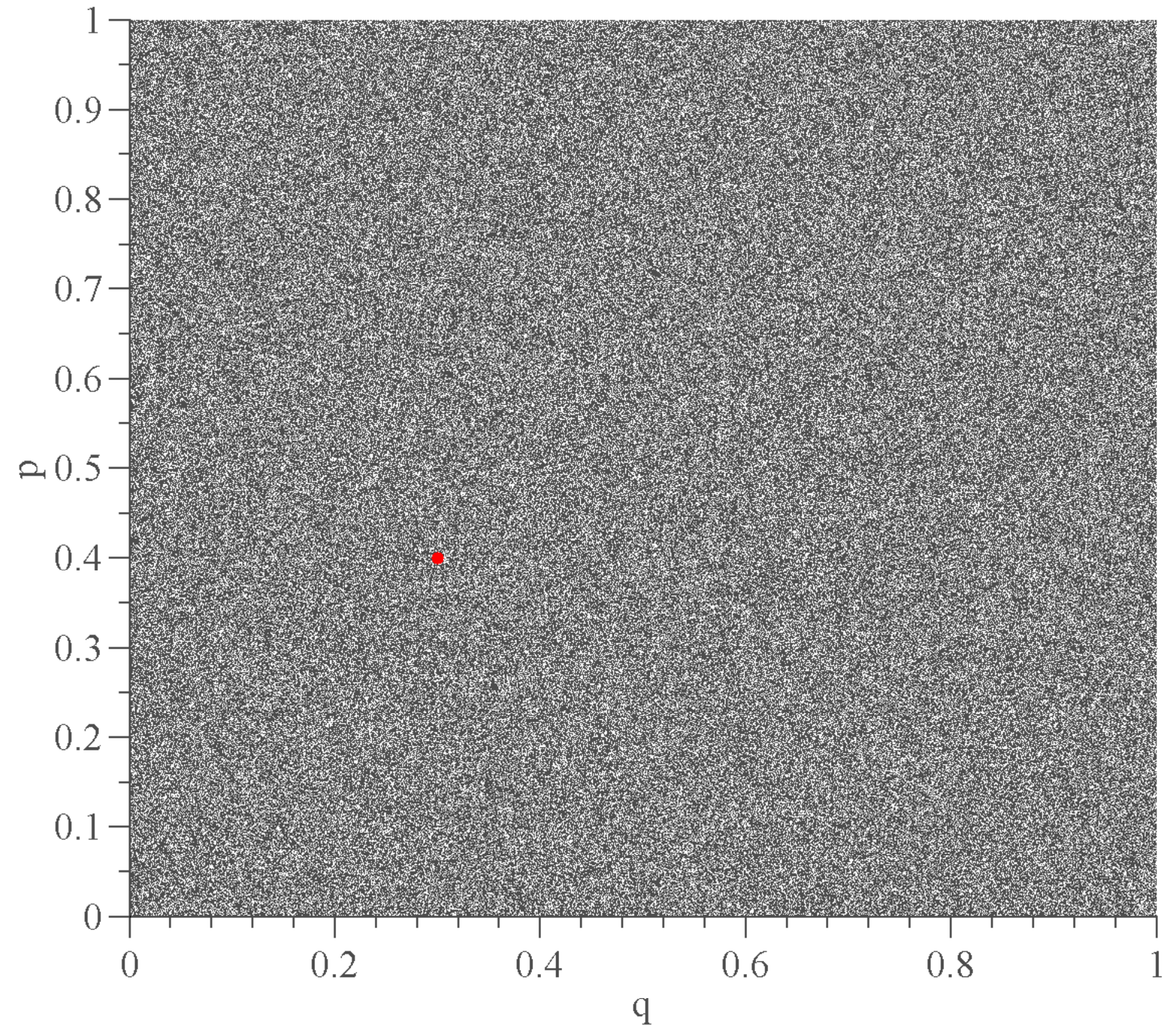}
\caption{\label{fig:one} Uniform filling of the phase-space by the Arnold's cat map suggesting the ergodicity of the mapping. The red dot indicates the initial conditions (0.3,0.4). 1 million mapping operations have been used to plot the figure.}
\end{figure}

Over the years, several maps have been proposed in the literature - Baker's map, Arnold's cat map, Duffing map, exponential map etc. However, according to our definition of time reversibility, most of the maps, including the extensively studied Baker's map \cite{hoover_book_12}, defined through the relation:
\begin{equation}
\begin{array}{ccl}

\left(q,p\right) & \xrightarrow{\mathcal{M}_{\text{Baker}}} & (\lambda_\alpha q, p/\alpha) \ \ \ \forall p < \alpha \\
& \xrightarrow{\mathcal{M}_{\text{Baker}}} & \left( 1 - \lambda_\beta + \lambda_\beta q, \left[ p - \alpha\right]/b \right) \ \ \ \forall p > \alpha
\end{array}
\label{eq:Baker}
\end{equation}
with $\beta = 1-\alpha$ and $\lambda_\alpha + \lambda_\beta \leq 1$, and the Arnold's cat map \cite{isaeva_06}, defined through the relation:
\begin{equation}
(q,p) \xrightarrow{\mathcal{M}_{\text{Arnold}}} (2q+p,p+q) \mod 1
\label{eq:Arnold}
\end{equation}
are not time reversible. {It must, however, be noted that the rotated Baker's map \cite{hoover2015simulation} follows our time-reversibility criteria.} While time reversibility of the maps is a sticky subject, the requirement of ergodicity is satisfied by several of the maps. For example take the Arnold's cat map beginning at the initial conditions $(q,p) = (0.3,0.4)$. The map quickly fills out the entire phase-space uniformly as shown in figure \ref{fig:one}. A similar behavior can be observed for the Baker's map.

Let us now look at the third desirable property of the maps that is exclusive to the nonequilibrium systems - dissipation. For a dissipative map, the Lyapunov exponents must sum up to being negative. The Baker's map, with appropriate selection of $\alpha, \lambda_\alpha$ and $\lambda_\beta$ allows dissipation \cite{rondoni_96}. However, the dissipation is zero for the Arnold's cat map \cite{isaeva_06} with $L_1 = \log[(3 + \sqrt{5})/2] = -L_2$. Thus, we see that most of the ``well studied'' maps do not satisfy all the three \textit{important} characteristics necessary for accurately representing the features of a dynamical system. 

\begin{figure}
\includegraphics[scale=0.28]{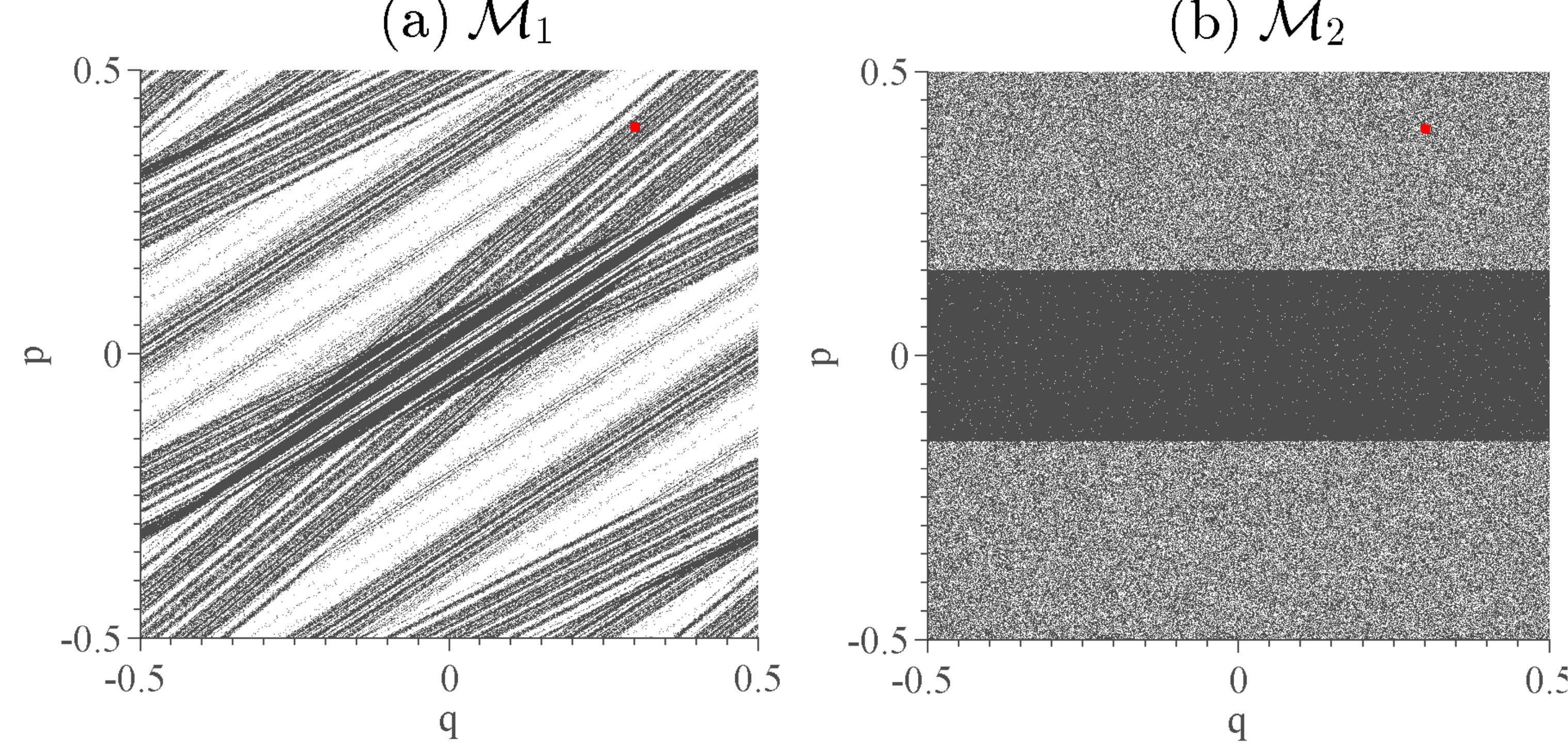}
\caption{\label{fig:two} The dissipative maps (a) $\mathcal{M}_1$, and (b) $\mathcal{M}_2$ for the initial conditions $(q,p,m) = (0.3,0.4,0.15)$ obtained through 1 million operations.}
\end{figure}

In 1996, a new map of a square [$(-0.5,-0.5) \times (0.5,0.5)$] onto itself has been proposed \cite{hoover_96} that satisfies all three requirements. It comprises of piecewise linear operations which are symmetrical combinations of simple shear operations:
\begin{equation}
(q,p) \xrightarrow{\mathcal{M}_{Q}} (q + p,p), \ \ (q,p) \xrightarrow{\mathcal{M}_{P}} (q,q + p), \ \ 
\label{eq:simple_shear}
\end{equation}
and reflection operation $\mathcal{M}_R$, wherein an imaginary mirror (located at $\pm m \leq \pm 1/4$ from origin) maps proportionally the points lying towards its left to right, and vice-versa, according to the relation: 
\begin{equation}
\mathcal{M}_R \equiv \dfrac{x_{\text{right}} - m}{x_{\text{left}} - m} =\dfrac{2m-1}{2m}.
\label{eq:reflection}
\end{equation}
In (\ref{eq:reflection}), $x$ denotes the variables $q$ and $p$. Consolidated maps that comprise of symmetric combination of these three operations, such as:
\begin{equation}
\begin{array}{rcl}
\mathcal{M}_1 = \mathcal{M}_Q \mathcal{M}_P \mathcal{M}_R \mathcal{M}_P \mathcal{M}_Q, \\
\mathcal{M}_2 = \mathcal{M}_Q \mathcal{M}_R \mathcal{M}_P \mathcal{M}_R \mathcal{M}_Q \\
\end{array}
\label{eq:combo}
\end{equation}
are ergodic, time-reversible as well as have provisions for allowing dissipation (depending on the value of $m$). Two such cases due to the mappings $\mathcal{M}_1$ and $\mathcal{M}_2$ with initial conditions $(q,p,m)=(0.3,0.4,0.15)$ are shown in figure \ref{fig:two}.
The multifractal nature of the dynamics is evident in figure \ref{fig:two}. Unfortunately, however, the simple mappings shown in (\ref{eq:combo}) are unable to match the complex multifractal nature of the continuous dynamical systems.

Recently, the Ian Snook memorial 2015 challenge has been posted \cite{ian_snook_15} regarding the development of new maps that are more complicated than the ones highlighted in figure \ref{fig:two}, while satisfying the three properties of maps highlighted before. The objective of this work is to develop new time-reversible ergodic maps with a parameter-dependent dissipation having complicated multifractal dynamics. We replace the three mapping operations shown in (\ref{eq:simple_shear}) and (\ref{eq:reflection}) with sinusoidal shears, and radial compression/expansion. The resulting linear combinations of the maps satisfy all targets that we set out to achieve. 

\section{Time-Reversible Ergodic Maps}
Amongst all the properties that we seek, time reversibility is the most challenging. So, let us first develop time-reversible maps. Consider an operation that maps a point $(q,p) \to (q^\prime,p)$. For the map to be time-reversible, the map must be an odd function of $p$. The simple shear operations shown in (\ref{eq:simple_shear}) constitute one such way. Let us look at another such approach where the mapping function is given by:
\begin{equation}
(q,p) \xrightarrow{\mathcal{M}_Q} (q+\sin p,p).
\label{eq:my_Q}
\end{equation} 
If $\mathcal{T}$ is the time-reverse operator that maps $p \to -p$, then the equation represented by \ref{eq:my_Q} is time-reversible:
\begin{equation}
(q,p) \xrightarrow{\mathcal{M}_Q} (q+\sin p,p) \xrightarrow{\mathcal{T}} (q+\sin p,-p) \xrightarrow{\mathcal{M}_Q} (q,-p).
\label{eq:time_reversible_my_Q}
\end{equation} 
In essense $\mathcal{T} \mathcal{M}_Q \mathcal{T} \mathcal{M}_Q \left\lbrace q,p \right\rbrace = \left\lbrace q,p \right\rbrace $. It is not difficult to see that $\mathcal{T}$ and its inverse, $\mathcal{T}^{-1}$, are additive inverse. In physically relevant terms, the mapping highlighted in \ref{eq:my_Q} consitutes the case of sinusoidally driven shear. Likewise let us define the mapping function for $p$:
\begin{equation}
(q,p) \xrightarrow{\mathcal{M}_P} (q,p+\sin q).
\label{eq:my_P}
\end{equation} 
To ensure that the map remains confined within the unit square, {we impose periodic boundary conditions by confining the variables within the unit square: $-0.5 \le q,q^\prime \le 0.5,-0.5 \le p,p^\prime \le 0.5$}. Any symmetric combination of $\mathcal{M}_Q$ and $\mathcal{M}_P$ results in a time-reversible mapping. Let us take the simple case of the mapping $\mathcal{M}_1 = \mathcal{M}_Q\mathcal{M}_P\mathcal{M}_Q$, and prove its time reversibility:
\begin{equation}
\begin{array}{l}
\mathcal{T} \left(\mathcal{M}_Q\mathcal{M}_P\mathcal{M}_Q \right) \mathcal{T} \left(\mathcal{M}_Q\mathcal{M}_P\mathcal{M}_Q \right) \left\lbrace q,p \right\rbrace \\

=\mathcal{T} \mathcal{M}_Q\mathcal{M}_P \left(\mathcal{M}_Q  \mathcal{T} \mathcal{M}_Q \right) \mathcal{M}_P\mathcal{M}_Q  \left\lbrace q,p \right\rbrace \\

= \mathcal{T} \mathcal{M}_Q \left(\mathcal{M}_P \mathcal{T}^{-1}  \mathcal{M}_P \right) \mathcal{M}_Q \left\lbrace q,p \right\rbrace \\

= -\mathcal{T} \mathcal{M}_Q \left(\mathcal{M}_P \mathcal{T}  \mathcal{M}_P \right) \mathcal{M}_Q \left\lbrace q,p \right\rbrace \\

= -\mathcal{T} \left(\mathcal{M}_Q \mathcal{T}^{-1}  \mathcal{M}_Q \right) \left\lbrace q,p \right\rbrace \\

= \left\lbrace q,p \right\rbrace
\end{array}
\label{eq:reversibility_proof}
\end{equation} 
Similarly, the more complicated mappings $\mathcal{M}_2 = \mathcal{M}_Q\mathcal{M}_P\mathcal{M}_Q\mathcal{M}_P\mathcal{M}_Q$ and $\mathcal{M}_3 = \mathcal{M}_P\mathcal{M}_Q\mathcal{M}_Q\mathcal{M}_P$ can be shown to be time reversible as well. However, the asymmetric cases like: $\mathcal{M}=\mathcal{M}_Q\mathcal{M}_P$ are not time reversible. 

\begin{figure*}
\includegraphics[scale=0.45]{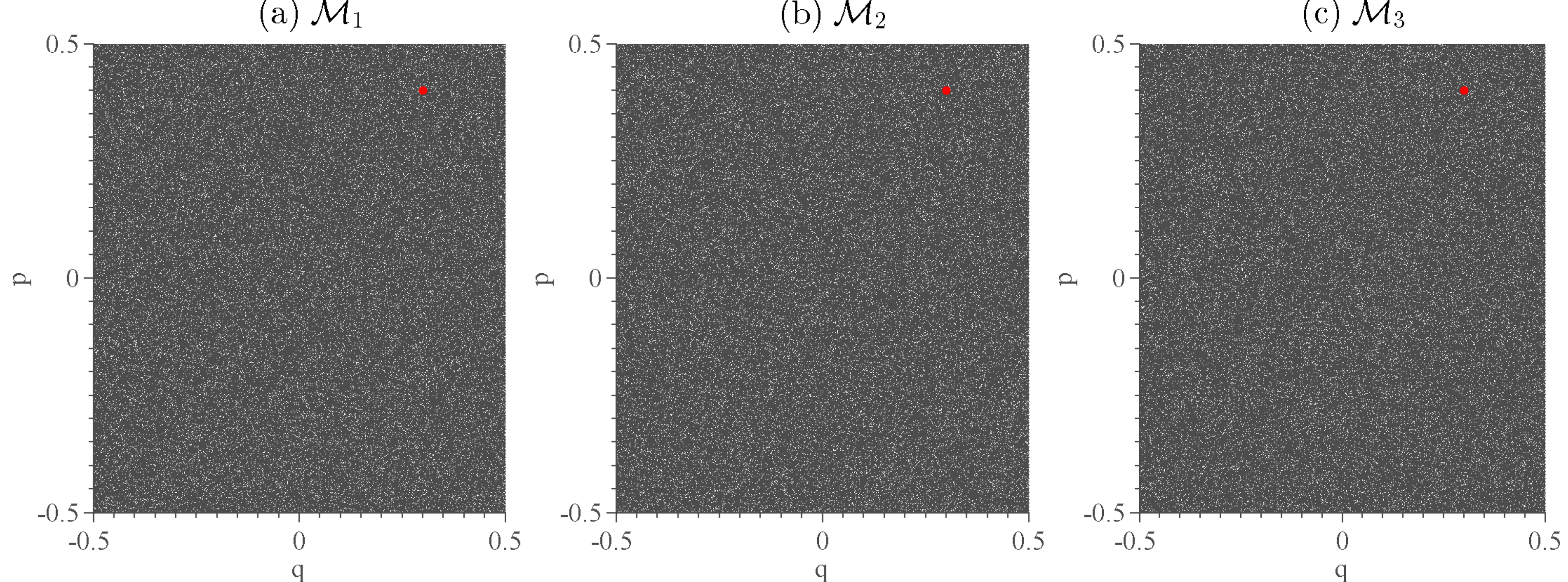}
\caption{\label{fig:three} Time-reversible ergodic maps initiating from the red dot ($(q,p) = (0.3,0.4)$) based upon sinusoidally driven shear operations: (a) corresponds to $\mathcal{M}_1 = \mathcal{M}_Q \mathcal{M}_P \mathcal{M}_Q$, (b) corresponds to $\mathcal{M}_2 = \mathcal{M}_Q \mathcal{M}_P \mathcal{M}_Q  \mathcal{M}_P \mathcal{M}_Q$, and (c) corresponds to $\mathcal{M}_3 = \mathcal{M}_P \mathcal{M}_Q \mathcal{M}_Q \mathcal{M}_P$. The results are plotted for 1 million operations. Notice that the mapping operations fill the entire square uniformly, suggesting that the maps are ergodic.}
\end{figure*}

Let us now investigate numerically the ergodic properties of the proposed $\mathcal{M}_1$, $\mathcal{M}_2$ and $\mathcal{M}_3$. By definition, we call a map ergodic if it comes arbitrary closer to all possible points, which in this case represents all possible points lying within the unit square $[-0.5,-0.5] \times [0.5,0.5]$. Thus, for the maps to be ergodic, they must sample the points according to a uniform distribution for both the $q$ and $p$ variables. Phase-potrait post 1 million mapping operations corresponding to the three maps  are shown in figure \ref{fig:three}. For each case, the mapping starts from the initial condition: $(q,p) \equiv (0.3,0.4)$. The resulting maps indicate a uniform coverage within in the unit square, suggesting that the maps are ergodic. However, mere visual inspection is not enough.

\subsection{Statistical Independence of the variables $q$ and $p$}
\begin{figure}
\includegraphics[scale=0.40]{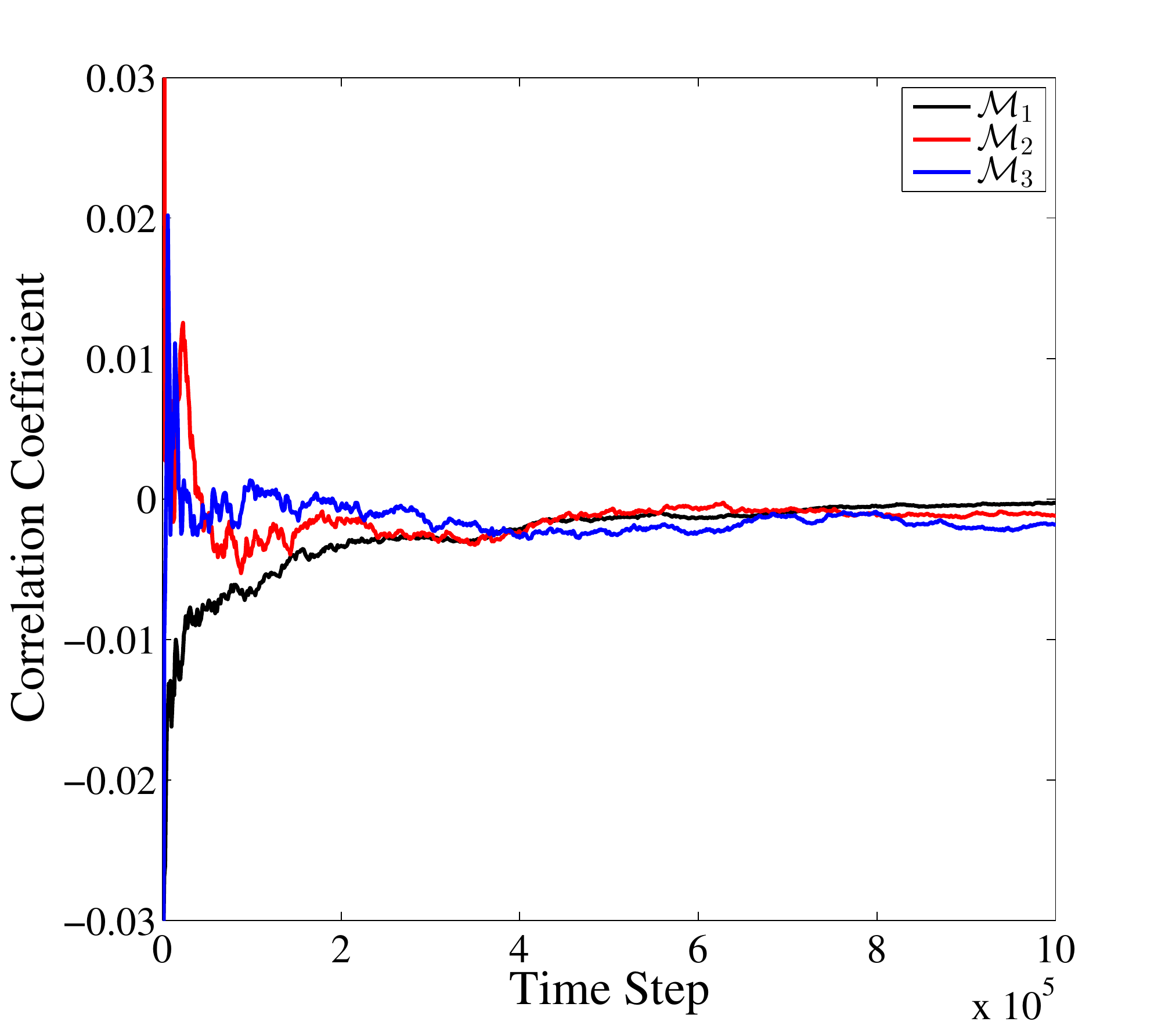}
\caption{\label{fig:correl_coefficient_evolution} Temporal evolution of correlation coefficient, calculated using (\ref{eq:correlation_coefficient}), for the three maps. The figure suggests that the correlation coefficient has converged.}
\end{figure}

To check for ergodicity, we need to show that the maps sample, for both $q$ and $p$ uniformly. Additionally, {the sequence of $q_i$ must be statistically independent from the sequence $p_i$ as well as from its subsequent realizations $q_{i+k}$. In other words, the sampling must be such that the values of $q$ and $p$ have zero correlation with each other, and $\delta-$ autocorrelation with itself. The correlation coefficient between the sequence $q$ and $p$ is defined as:
\begin{equation}
\rho_{q,p} = \dfrac{E \left[ qp \right] - E \left[ q \right] E \left[ p \right]}{\sigma_q \sigma_p} = \dfrac{\dfrac{\sum\limits_{i=1}^N q_ip_i}{N} - \dfrac{\sum\limits_{i=1}^N q_i }{N} \times \dfrac{\sum\limits_{i=1}^N p_i}{N}}{\sigma_q \sigma_p},
\label{eq:correlation_coefficient}
\end{equation}}
{where $E\left[ q \right]$ and $E\left[ p \right]$ are the expected values of $q$ and $p$, $E\left[ qp \right]$ is the expected value of the product $qp$, and $\sigma_q$ and $\sigma_p$ are the standard deviations of $q$ and $p$, respectively. $N$ represents the total number of samples. The correlations between $q$ and $p$ for the three maps are shown in the table \ref{tab:table1}, and their temporal evolutions are shown in figure \ref{fig:correl_coefficient_evolution}.} 
\begin{table}
\caption{Correlations between $q$ and $p$ for the three maps.}
\label{tab:table1}
\begin{tabular*}{0.275\textwidth}{@{\extracolsep{\fill}}cc@{}}
\hline \hline
 & Correlation Coefficient\\ \hline  \hline
$\mathcal{M}_1$ & -0.0003 \\ 
$\mathcal{M}_2$ & -0.0012 \\
$\mathcal{M}_3$ & -0.0019 \\ \hline
\end{tabular*}
\end{table}
{The figure suggests that the correlation coefficient has converged. It is evident that the variables have statistically insignificant correlations for all three maps.}

\begin{table*}[]
\centering
\caption{The comparison of joint moments: true joint moment gives the theoretical joint moment if the variables are identically and independently uniformly distributed, $E[q^mp^n]$ gives the joint moment obtained numerically, and $E[q^m]E[q^n]$ gives the product of marginal moments obtained numerically. Notice that for each row, the true moment matches closely with the numerically computed joint moments and product of marginal moments. The equality suggests that the variables $q$ and $p$ are statistically independent.}
\label{tab:hom_ind_test}
\begin{tabular*}{0.95\textwidth}{@{\extracolsep{\fill}}|c|c|cc|cc|cc|@{}}
\hline
\multirow{2}{*}{$(m,n)$} & \multirow{2}{*}{True Moment} & \multicolumn{2}{c}{$\mathcal{M}_1$} & \multicolumn{2}{c}{$\mathcal{M}_2$} & \multicolumn{2}{c}{$\mathcal{M}_3$} \\
                         &                              & $E[q^mp^n]$     & $E[q^m]E[p^n]$    & $E[q^mp^n]$     & $E[q^m]E[p^n]$    & $E[q^mp^n]$     & $E[q^m]E[p^n]$    \\ \hline \hline
(2, 2)                   & 0                            & -0.00002302     & 0.00000000        & -0.00009769     & 0.00000000        & -0.00015634     & 0.00000000        \\
(3, 3)                   & 0.00694444                   & 0.00694239      & 0.00694168        & 0.00692949      & 0.00692791        & 0.00696127      & 0.00695927        \\
(4, 4)                   & 0                            & 0.00000000      & 0.00000000        & -0.00000168     & 0.00000000        & -0.00000464     & 0.00000000        \\
(5, 5)                   & 0.00015625                   & 0.00015608      & 0.00015611        & 0.00015608      & 0.00015581        & 0.00015680      & 0.00015661        \\
(6, 6)                   & 0                            & 0.00000000      & 0.00000000        & -0.00000007     & 0.00000000        & -0.00000020     & 0.00000000        \\
(7, 7)                   & 0.00000498                   & 0.00000497      & 0.00000498        & 0.00000499      & 0.00000497        & 0.00000501      & 0.00000499        \\
(8, 8)                   & 0                            & 0.00000000      & 0.00000000        & 0.00000000      & 0.00000000        & -0.00000001     & 0.00000000        \\ \hline
\end{tabular*}
\end{table*}

{Ideally, if the variables $q$ and $p$ are statistically independent, the following relation holds true for all values of $m$ and $n$:
\begin{equation}
E[q^mp^n] = E[q^m]E[p^n]
\label{eq:higher_order_moment_independence}
\end{equation}
The test of statistical independence based on (\ref{eq:correlation_coefficient}) is a special case of (\ref{eq:higher_order_moment_independence}), where $m=n=1$. We, therefore, now compare the joint higher order moments with the product of equivalent marginal moments. The results are shown in table \ref{tab:hom_ind_test}. The theoretical moments have been computed (with the assumption that the variables are independent) through the relation:}  
\begin{equation}
E[q^mp^n] = \dfrac{0.5^{m+1}-(-0.5)^{m+1}}{m+1} \times \dfrac{0.5^{n+1}-(-0.5)^{n+1}}{n+1}
\label{eq:theoretical_moment}
\end{equation}
{The true moment matches closely with the numerically computed joint moments and product of marginal moments. The equality suggests that the variables $q$ and $p$ are statistically independent.}

{Now let us look at the independence of the variables from their subsequent realizations using autocorrelation function. The autocorrelation $\zeta(k)$ measures the correlation of a sequence with itself at different points in time, and can be calculated as:
\begin{equation}
\zeta_{q,q}(k) = \dfrac{\sum\limits_{i=1}^{N-k} \left(q_i -E[q] \right) \left( q_{i+k} - E[q] \right) }{\sigma_q^2},
\label{eq:autocorrelation}
\end{equation}}

\begin{figure}
\includegraphics[scale=0.50]{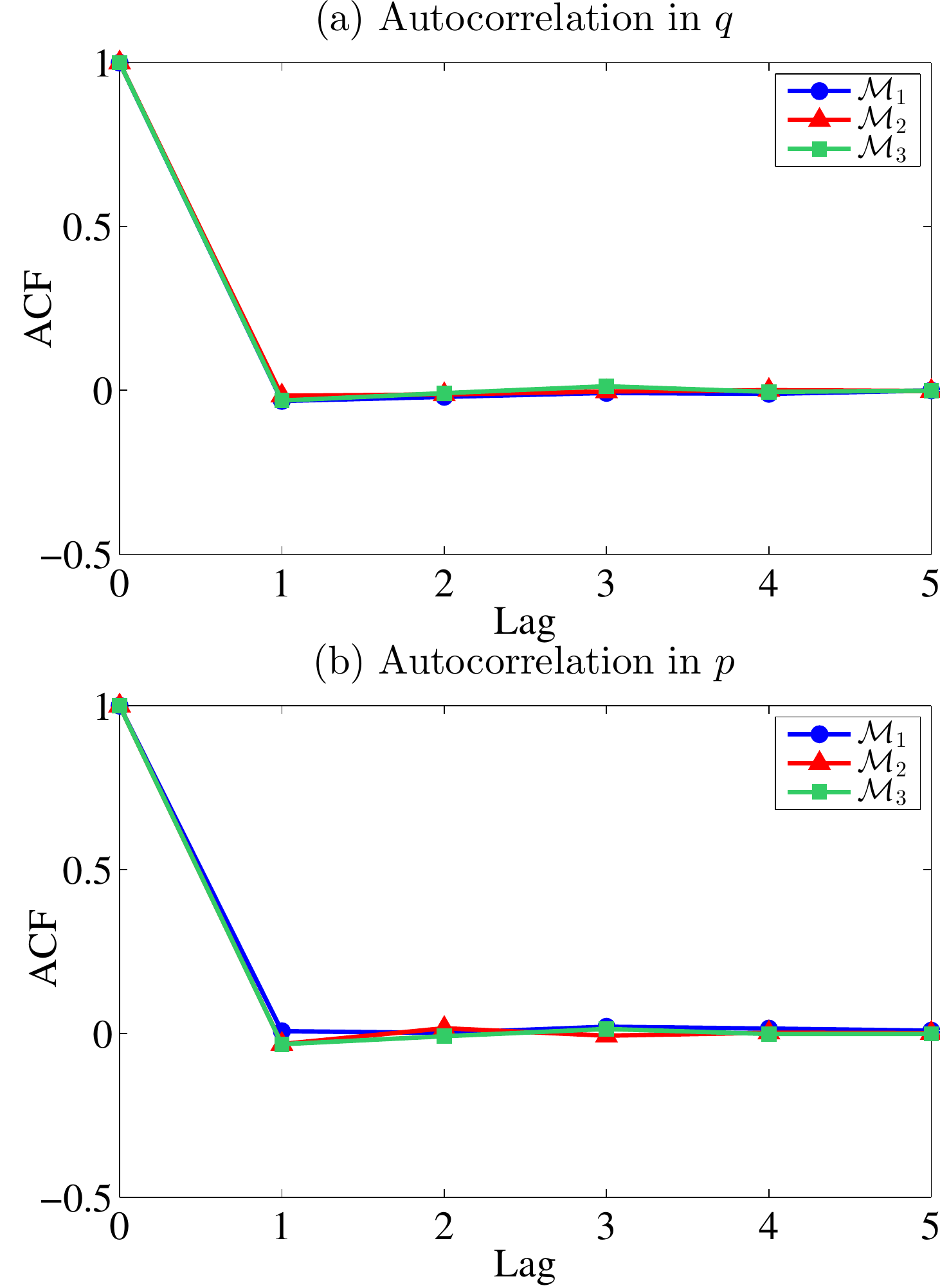}
\caption{\label{fig:four} The autocorrelation functions (ACF) due three maps: (a) for the variable $q$, and (b) for the variable $p$. All six variables are delta-correlated, implying that the value obtained in $(i+1)^{\text{th}}$ iteration is independent of the $i^{\text{th}}$ iteration.}
\end{figure}

The autocorrelation function (ACF) of $q$ and $p$ due to the three maps are shown in figure \ref{fig:four}. {The figures reveal that there is an autocorrelation of the order of 0.01 for each of the variables, which may be taken as small enough to be neglected. The finite value of autocorrelation, however, makes it difficult to use the maps for generating uniform random numbers. For the remainder of this work, we treat the autocorrelation of the variables small enough to be neglected.}

\subsection{Tests for Uniformity}
The uniformity of the variables have been ascertained using three basic statistical tests -- the mean test, the variance test and the bucket test, as discussed next. 

\textbf{Mean Test}: The mean test checks if the average of a large number of realizations of the variables agrees with the theoretical value \cite{riley_book}. The theoretical average for a uniform random number distributed between $[-0.5,0.5$ is 0. The averages of the six variables obtained due to the three maps is shown in the table \ref{tab:table2}.
\begin{table}
\caption{$\langle q \rangle$ and $\langle p \rangle$ values for the three different maps.}
\label{tab:table2}
\begin{tabular*}{0.275\textwidth}{@{\extracolsep{\fill}}ccc@{}}
\hline \hline
 & $\langle q \rangle$ & $\langle p \rangle$ \\ \hline  \hline
$\mathcal{M}_1$ & $-3.3 \times 10^{-4}$ & $-1.3 \times 10^{-4}$ \\ 
$\mathcal{M}_2$ & $-3.2 \times 10^{-4}$ & $-1.4 \times 10^{-4}$ \\
$\mathcal{M}_3$ & $-1.6 \times 10^{-4}$ & $-4.63 \times 10^{-4}$ \\ \hline
\end{tabular*}
\end{table}
Performing a hypothesis test at 95\% confidence level reveals that the true mean lies between $\pm 5.65 \times 10^{-4}$. Thus, from the mean test, there is no statistical evidence to suggest that the variables $q$ and $p$ are not uniform.

\textbf{Variance Test}: Next we perform the variance test \cite{riley_book} where we compare the sample variance vs. the true variance (=1/12). The variances are listed in the table \ref{tab:table3}.
\begin{table}
\caption{Variances $\langle q^2 \rangle$ and $\langle p^2 \rangle$ values for the three different maps.}
\label{tab:table3}
\begin{tabular*}{0.275\textwidth}{@{\extracolsep{\fill}}ccc@{}}
\hline \hline
 & $\langle q^2 \rangle$ & $\langle p^2 \rangle$ \\ \hline  \hline
$\mathcal{M}_1$ & $0.0833$ & $0.0834$ \\ 
$\mathcal{M}_2$ & $0.0833$ & $0.0832$ \\
$\mathcal{M}_3$ & $0.0833$ & $0.0833$ \\ \hline
\end{tabular*}
\end{table}
The 95\% confidence interval of the true variance is $(0.0831,0.0836)$. Thus, we again see that there is no sufficient statistical evidence to suggest that the variables are not uniformly distributed.

\textbf{Bucket Test}: The entire data (of 1 million points) corresponding to each variable is grouped together in 100 bins. We then perform a chi-squared test \cite{riley_book} by computing the test statistic:
\begin{equation}
\chi^2 = \sum\limits_{i=1}^{100}\dfrac{\left( O_i - E_i\right)^2}{E_i}.
\end{equation}
Here $O_i$ and $E_i$ give the observed and expected counts in the $i^{\text{the}}$ bin The test statistic for the six variables are shown in table \ref{tab:table4}
\begin{table}
\caption{$\chi^2$ test statistic for the six ``random'' variables .}
\label{tab:table4}
\begin{tabular*}{0.275\textwidth}{@{\extracolsep{\fill}}ccc@{}}
\hline \hline
 & $q$ & $p$ \\ \hline  \hline
$\mathcal{M}_1$ & 96.005 & 81.447 \\ 
$\mathcal{M}_2$ & 98.904 & 82.015 \\
$\mathcal{M}_3$ & 120.804 & 99.130 \\ \hline
\end{tabular*}
\end{table}
The critical $\chi^2$ value corresponding to a probability of 0.05 with 99 degrees of freedom is 124.34. Since all values shown in table \ref{tab:table4} are lesser than it, we can say that there is a sufficient evidence that all variables are uniformly distributed.

All three tests (the mean test, the variance test and the bucket test) suggest that there is sufficient evidence that the random variables $q$ and $p$, for all three maps, are uniformly distributed between [-0.5,0.5]. As the two variables corresponding to a particular map are statistically independent as well, we therefore conclude that the maps sample uniformly from the unit square, and hence ergodic.

{Unfortunately, the autocorrelation of the variables, howsoever small, renders it difficult to use the maps as pseudo-random number generators. Additionally, all three maps fail the stringent ``Diehard Battery'' of tests for randomness \cite{marsaglia1996diehard}, which confirms that the random numbers generated using the three maps cannot be used as pseudo-random number generators.} 

\subsection{Lyapunov Exponents} 
\begin{figure}
\includegraphics[scale=0.35]{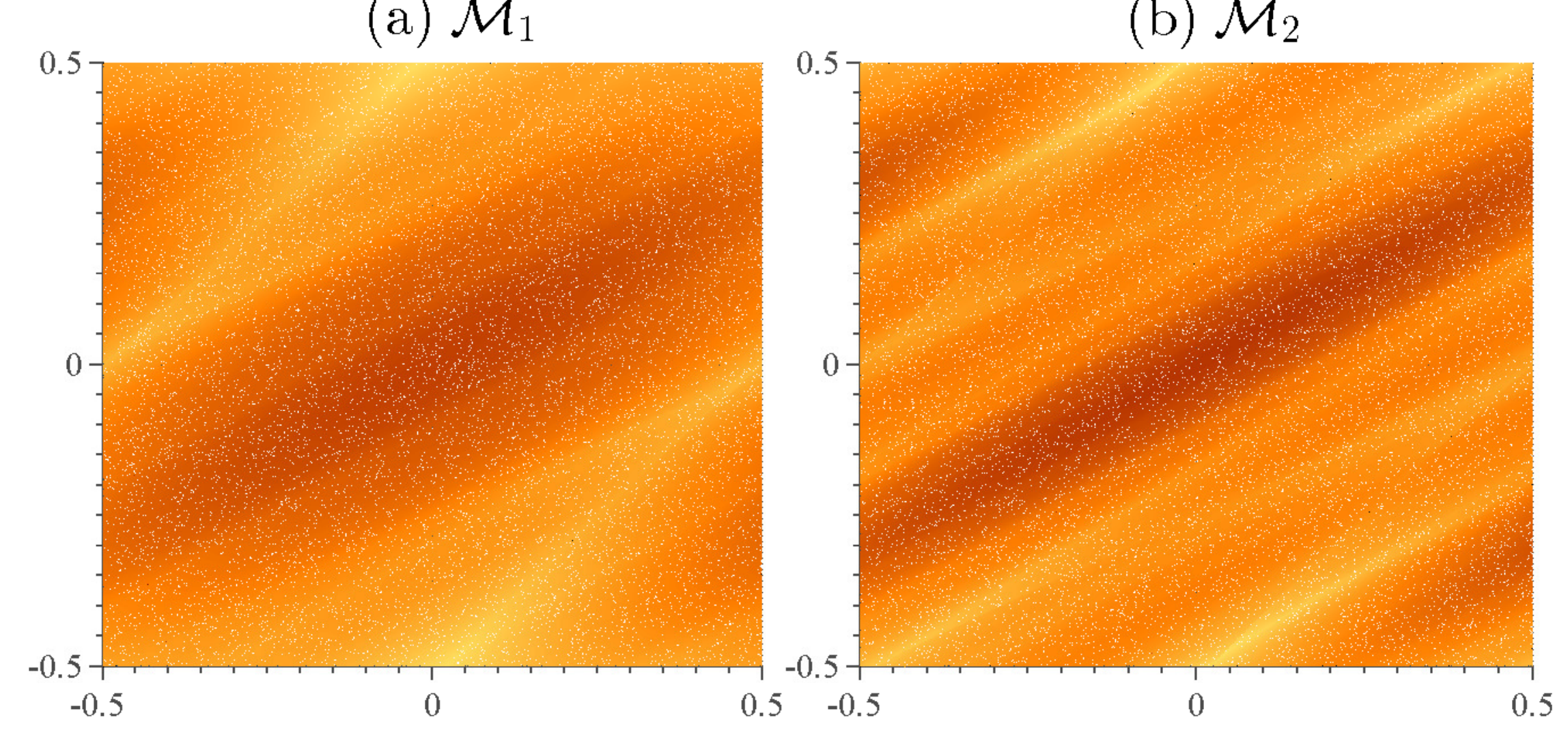}
\caption{\label{fig:five} Local Lyapunov exponents for the two maps obtained using 1 million iterations. The darker the color, the larger is the value of the exponent. The exact color correspondence with values is different for each figure, however. Notice that the distribution of lyapunov exponents is far from being uniform. Also, note that the values are never negative.}
\end{figure}

Dissipation invariably involves compressible phase-space flows, with the Lyapunov exponents summing up to be negative. The largest Lyapunov exponent, $L_1$, provides the exponential rate at which two nearby trajectories separate, while the sum of $L_1$ and $L_2$, the smallest Lyapunov exponent, provides the rate at which the area changes. The maps developed so far are area-preserving, and hence non dissipative. The two Lyapunov exponents corresponding to these maps are paired, and therefore sum up to zero. To obtain the two Lyapunov exponents, apart from the original map, two satellite maps (separated from the original by $\Delta = 0.000001$) are solved \cite{hoover_book_12}. After every iteration, Gram-Schmidt orthonormalization is performed, and the distance between the satellite values from the original values are normalized such that they are $\Delta$ distance apart. The two Lyapunov exponents for the maps are shown in table \ref{tab:table5}. It is evident that $L_1 + L_2 \approx 0$, confirming that the maps are area preserving, and thus, have no provisions for dissipation. 
\begin{table}
\caption{Lyapunov exponents $L_1$ and $L_2$ corresponding to the three maps.}
\label{tab:table5}
\begin{tabular*}{0.275\textwidth}{@{\extracolsep{\fill}}ccc@{}}
\hline \hline
 & $L_1$ & $L_2$ \\ \hline  \hline
$\mathcal{M}_1$ & $1.268_7$ & $-1.268_7$ \\ 
$\mathcal{M}_2$ & $2.207_4$ & $-2.207_1$ \\
$\mathcal{M}_3$ & $1.703_3$ & $-1.703_3$ \\ \hline
\end{tabular*}
\end{table}
It is interesting to note that despite providing a uniform measure and area conservation, the distribution of local Lyapunov exponents for these maps is far from being uniform, as shown in figure \ref{fig:five}. 

\section{Dissipative Time Reversible Maps}
In this section, we create a simple parameter dependent time-reversible dissipative map, and combine with the maps developed in the previous section to obtain a composite map that, depending on the parameter, allows dissipation. 

\begin{figure}
\includegraphics[scale=0.60]{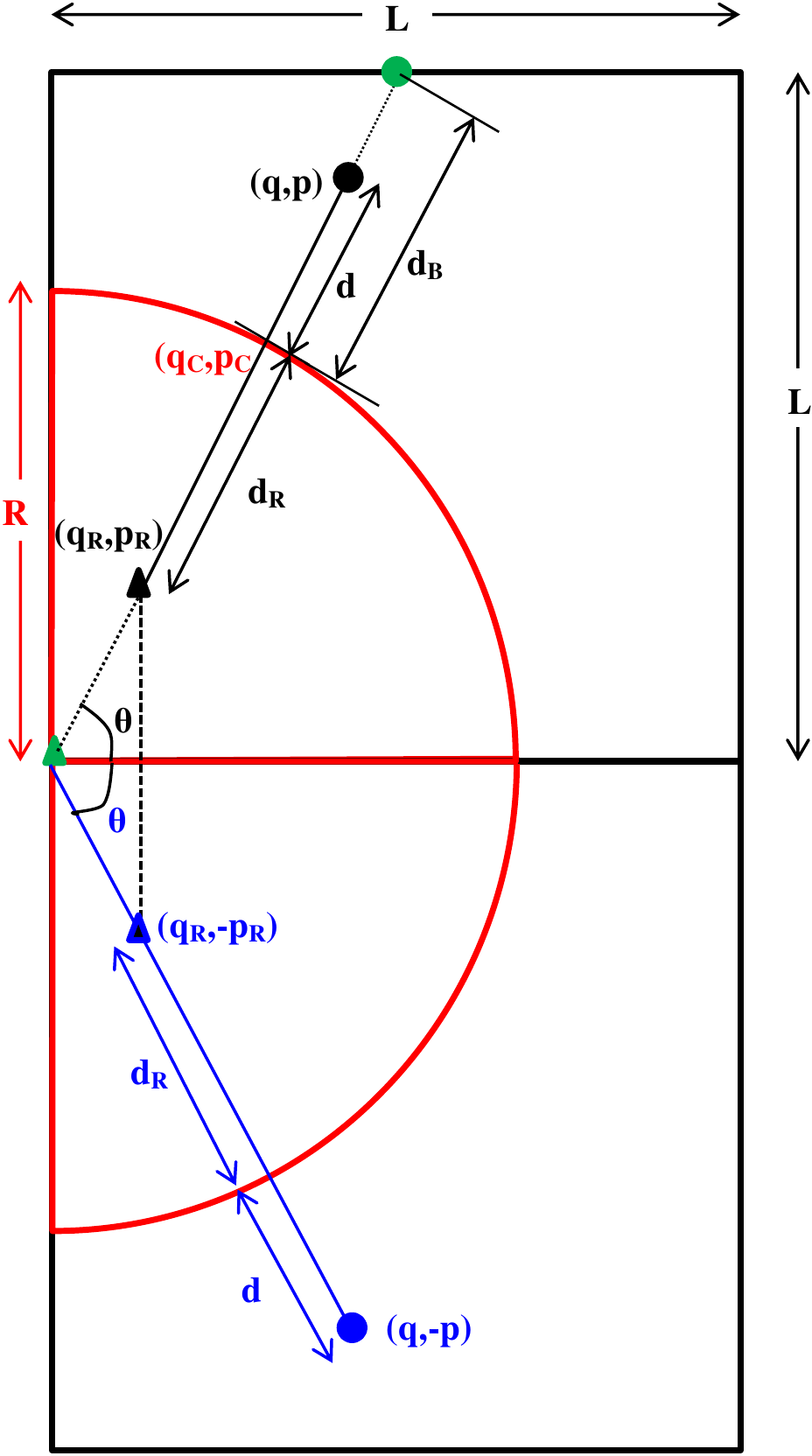}
\caption{\label{fig:six} Radial compression that maps a point lying outside the circle, $(q,p)$, to a point inside the circle $(q_R,p_R)$, and vice versa. Owing to the symmetry only the first quadrant is sufficient for the analysis. The compression proceeds such that the the green dot, lying at the boundary, is mapped to the origin (the green triangle). All remaining points are mapped proportionally. The blue colored (bottom square) points denotes the location of points post the time-reversal operator $\mathcal{T}$ has been applied. $\mathcal{T}$ operates on $(q_R,p_R)$ to take it to $(q_R,-p_R)$.  Since the angle post $\mathcal{T}$ remains unchanged, and so does the distance from the circle, the mappint $\mathcal{M}_R$ takes this point to $(q,-p)$, thus, proving time-reversibility. }
\end{figure}

\subsection{Dissipative Maps}
Consider the radial compression mapping $\mathcal{M}_R$, for the first quadrant, shown in figure \ref{fig:six}, where a point, $(q,p)$, that makes an angle $\theta$ is mapped to a point $(q_R,p_R)$ lying along the same line. If $d$ is the distance between the circle and the point, then post mapping, the updated location is at a distance $d_R$ from the circle. The updated coordinates are given by:
\begin{equation}
\begin{array}{rcl}
q_R = q_C - d_R \times \cos \theta & = & q_C - \dfrac{Rd}{d_B}\times \cos \theta \\
p_R = p_C - d_R \times \sin \theta & = & p_C - \dfrac{Rd}{d_B}\times \sin \theta \\
\end{array}
\label{eq:radial_compression_in_cicle}
\end{equation}
Likewise, a point lying within the circle is mapped outside of it through the relations:
\begin{equation}
\begin{array}{rcl}
q = q_C + d \times \cos \theta & = & q_C + \dfrac{d_Bd_R}{R}\times \cos \theta \\
p = p_C + d \times \sin \theta & = & p_C + \dfrac{d_Bd_R}{R}\times \sin \theta \\
\end{array}
\label{eq:radial_compression_out_circ}
\end{equation}
Similar relations can be obtained for other quadrants as well. Before proceeding further let us prove graphically the time-reversibility of the map $\mathcal{M}_R$ (see figure \ref{fig:six}). The time-reversal operator $\mathcal{T}$ acting upon this brings the point to the fourth quadrant $(q_R,-p_R)$. Since the magnitude of the angle $\theta$ that the point makes remains unchanged along with its distance from the circle, under $\mathcal{M}_R$, $(q_R,-p_R)$ gets mapped to $(q,-p)$, thereby satisfying time-reversibility. 

Depending upon the radius, $R$, of the circle different dissipation regimes can be obtained. For example, if $R$ is near zero, a limit cycle is obtained, while if $R$ is moderately greater than zero, multifractals are obtained. This mapping used in conjunction with $\mathcal{M}_Q$ and $\mathcal{M}_P$, in symmetric combinations, can be used to develop time-reversible ergodic dissipative maps.

{\subsection{Multifractal phase space}
We next explore numerically the phase-space due to two such maps:
\begin{equation}
\begin{array}{c}
\mathcal{M}_{D,1} = \mathcal{M}_Q \mathcal{M}_P \mathcal{M}_R \mathcal{M}_P \mathcal{M}_Q,\\
\mathcal{M}_{D,2} = \mathcal{M}_Q \mathcal{M}_R \mathcal{M}_P \mathcal{M}_R \mathcal{M}_Q.
\end{array}
\end{equation}
For both of them we keep the initial conditions at $(q,p) = (0.3,0.4)$, and use two values of $R$: 0.20 and 0.30. These results are shown in figure \ref{fig:dissipative_phase_space}. It is evident that the resulting multifractal nature of the phase-space is significantly more complicated than the ones shown in the figure \ref{fig:two}.
\begin{figure}
\includegraphics[scale=0.35]{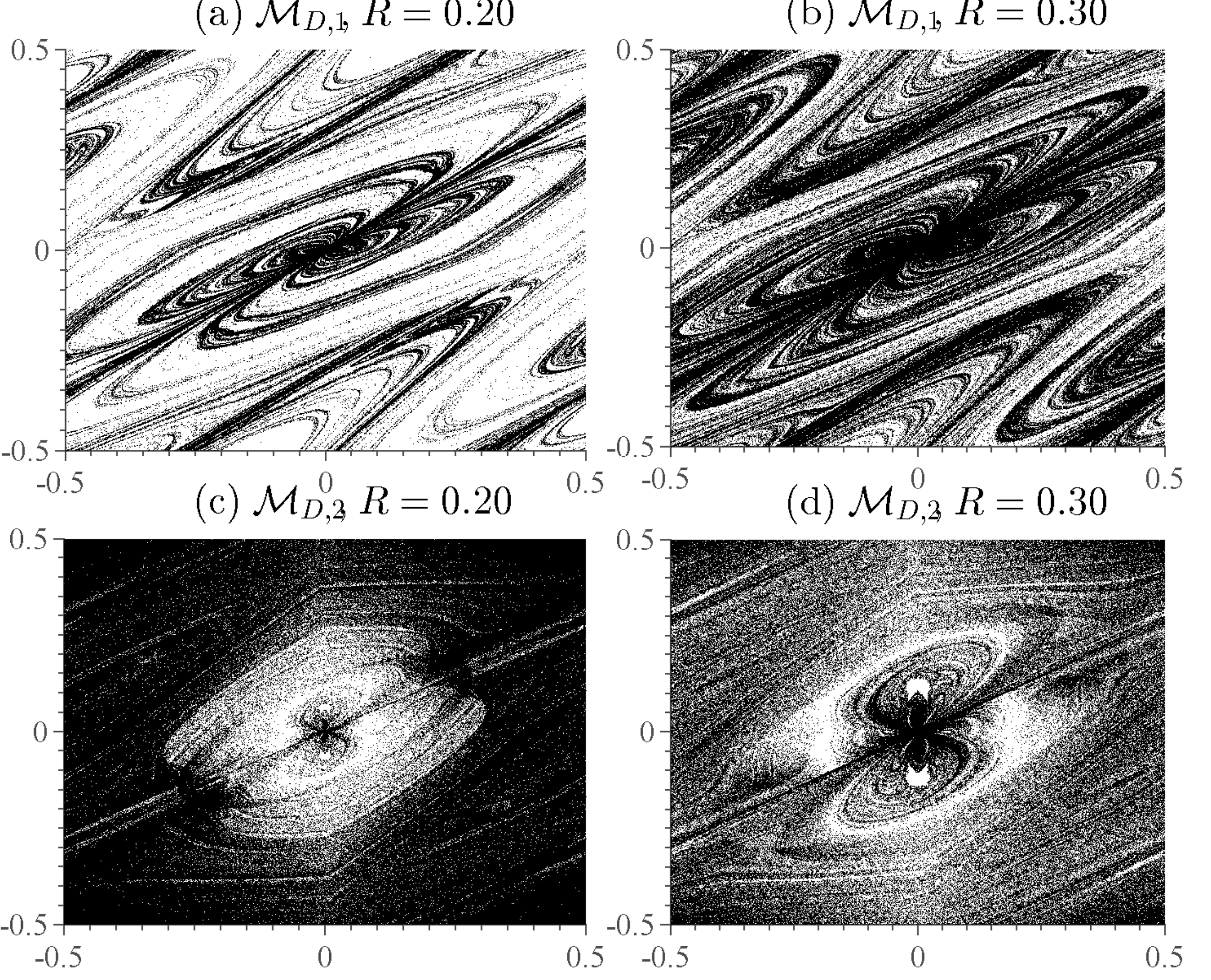}
\caption{\label{fig:dissipative_phase_space} The dissipative maps showing the local Lyapunov exponent corresponding to the two maps $\mathcal{M}_{D,1}$ and $\mathcal{M}_{D,2}$, and two different radii of the circle. The darker the color, the larger is the value of the local lyapunov exponent. Notice the radius dependence of the multifractal attractor. }
\end{figure}
There is a clear radius dependence of the phase-space: the multifractal nature of the phase space changes with radius. Let us now look at the ergodic properties of these maps. }

{\subsection{Ergodicity of dissipative maps}
One can ascertain the ergodicity of the dissipative maps with thousands of different initial conditions by (i) comparing the difference between the maximum and minimum of the resulting largest Lyapunov exponents, and (ii) comparing the equality of the phase-averages like $\langle q^n \rangle, \langle p^n \rangle$ and $\langle q^np^n \rangle$ with $n=1,2,3$ arising due to these initial conditions. If the difference between the maximum and minimum largest Lyapunov exponent and the phase-averages are small, the map (corresponding to a given $R$) is ergodic. On the other hand, if the differences are not small, the maps is not ergodic (corresponding to a given $R$). The ergodicity of the maps is tested using 2500 initial conditions obtained by dividing the unit square into a grid of $50 \times 50$ squares. Coordinate of each node serves as an initial condition. The map $\mathcal{M}_{D,1}$ is iterated for 1 million steps, while the map $\mathcal{M}_{D,2}$ is iterated for 5 million steps.}

\begin{figure}
\includegraphics[scale=0.45]{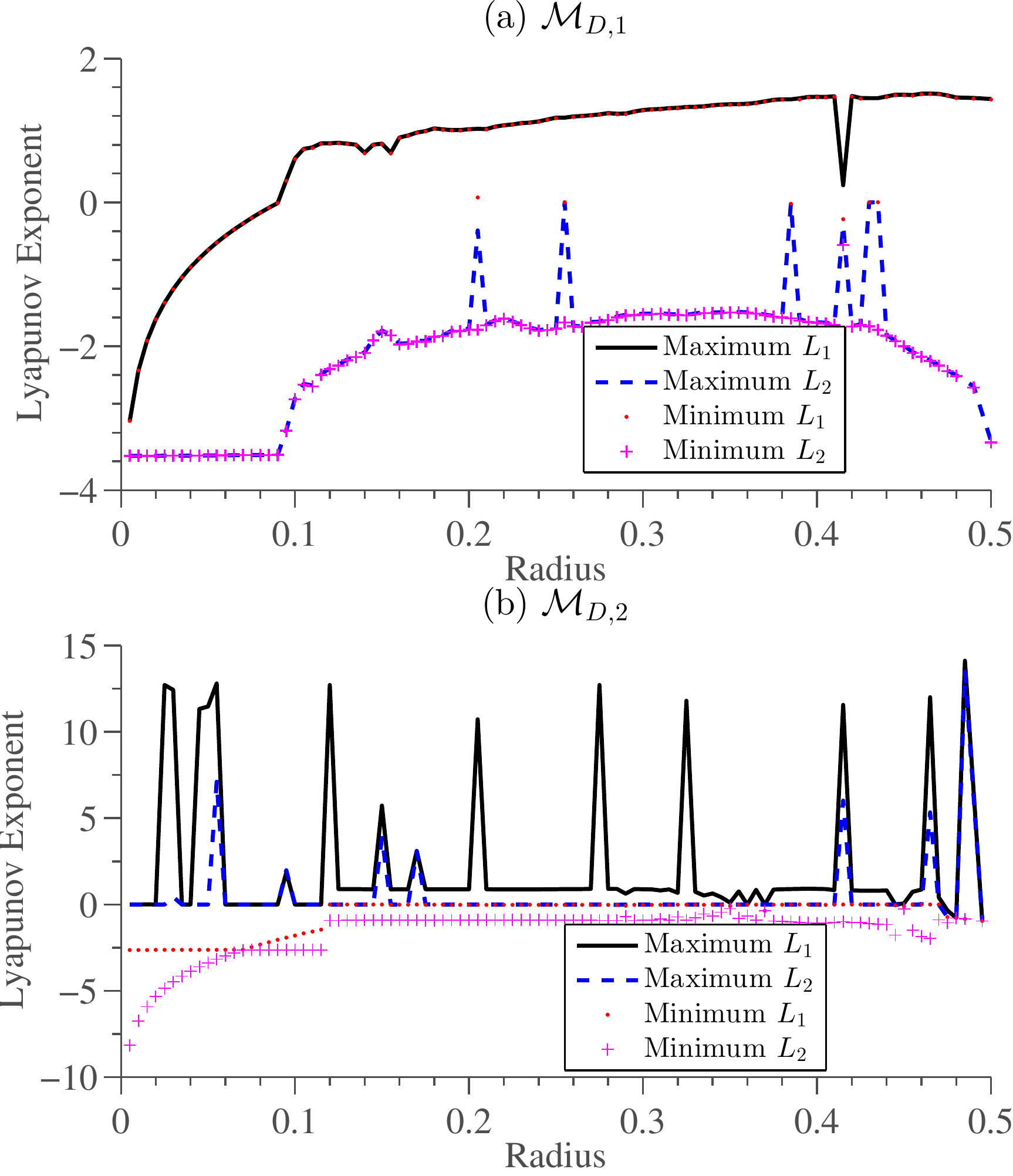}
\caption{\label{fig:lyap_ergodicity} The maximum and minimum largest Lyapunov exponent, $L_1$, and the smallest Lyapunov exponent, $L_2$, obtained from a grid of 2500 initial conditions. The maps are ergodic if the difference between the maximum and minimum Lyapunov exponents are comparable for a given radius. The results indicate that the map $\mathcal{M}_{D,1}$ is ergodic for the better part of the radii spectrum (except at the radii where there is a significant difference between the maximum and minimum Lyapunov exponents). The  map $\mathcal{M}_{D,2}$, on the other hand, shows a significant deviation between the maximum and minimum largest Lyapunov exponent for almost every radius, and therefore, is nonergodic.}
\end{figure}

{The maximum and the minimum largest and smallest Lyapunov exponents for the two maps (obtained from 2500 initial conditions) are shown in figure \ref{fig:lyap_ergodicity}. The results indicate that the map $\mathcal{M}_{D,1}$ is ergodic for the better part of the radii spectrum (except at the radii where there is a significant difference between the maximum and minimum Lyapunov exponents). The  map $\mathcal{M}_{D,2}$, on the other hand, shows a significant deviation between the maximum and minimum largest Lyapunov exponent for almost every radius, and therefore, is nonergodic. A similar conclusion can be drawn by looking at the different moments. It is interesting to note that regardless of map, the sum of the Lyapunov exponents is less than 0, suggesting that there is a finite amount of dissipation occurring. Hence, we come to the conclusion that only $\mathcal{M}_{D,2}$ satisfies ergodicity, time-reversibility and dissipation.}

{For $\mathcal{M}_{D,1}$, at small values of the radius, both $L_1$ and $L_2$ are negative, suggesting the presence of stable periodic orbits. As radius increases, $L_1$ becomes greater than zero, suggesting the presence of a chaotic regime. The sum of both Lyapunov exponents, however, remains negative. It is not very surprising that the sum has a minimum value (of around -0.09) near $R=0.385$. A simple calculation reveals that at $R=\sqrt{2/\pi} \times 0.5 \approx 0.39$, the area of the circle is exactly the same as the area of remaining square so that the dissipation is minimum (it cannot be zero owing to the local dependence of compression/expansion). Also evident is the fact that the Kaplan-Yorke dimension is smaller than the embedding dimension of 2 for all radii.}

\subsection{Fractals and local Lyapunov exponents}
\begin{figure}
\includegraphics[scale=0.325]{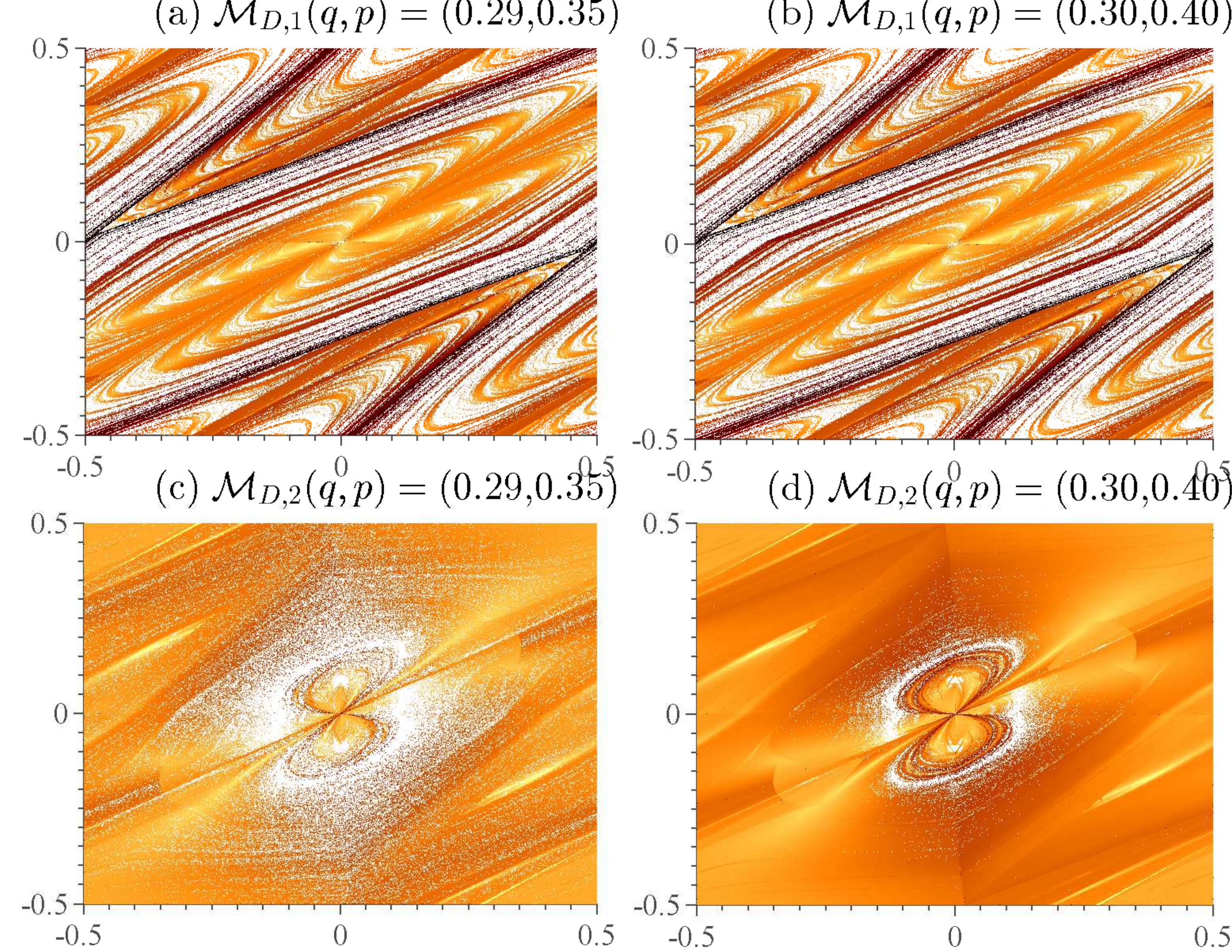}
\caption{\label{fig:seven} The dissipative maps showing the local Lyapunov exponent corresponding to the two maps $\mathcal{M}_{D,1}$ and $\mathcal{M}_{D,2}$, at $R=0.25$ for two different initial conditions. The darker the color, the larger is the value of the local lyapunov exponent. Notice the initial condition dependence of the multifractal attractor for $\mathcal{M}_{D,2}$. }
\end{figure}

{We now look at the local Lyapunov exponents and the strange attractors corresponding to the two maps. For this purpose, we keep $R = 0.25$, and use two different initial conditions: $(q,p) = (0.3,0.4)$ and (0.29,0.35). The results are shown in figure \ref{fig:seven}. Both the maps show a multifractal attractor, suggesting that these maps have one of the most important feature of nonequilibrium dynamical systems -- dissipation. However, the nature of the attractor differs for $\mathcal{M}_{D,2}$, depending upon the initial conditions chosen. For $\mathcal{M}_{D,1}$, on the other hand, initial conditions do not have a significant influence over the nature of the attractor.}

\section{Summary and Conclusions:}
In this work we have proposed two basic mapping operations - (i) time-reversible non-dissipative maps ($\mathcal{M}_P$ and $\mathcal{M}_Q$) inspired by sinusoidally driven shearing systems, and (ii) time-reversible dissipative map ($\mathcal{M}_R$) which radially compresses/expands the system. All symmetric combinations of these maps result in time-reversible mappings. However, not all of them are (i) dissipative and (ii) ergodic.

The time-reversible non-dissipative maps sample from a uniform distribution as evidenced from the three basic statistical tests performed in this study. However, due to the \textit{small} autocorrelation function, these maps cannot be utilized as a viable alternative to the pseudo-random number generators. Additionally, the maps fail the more stringent tests for randomness. 

Developed as a solution to the recently posted 2015 Ian Snook prize problem, the dissipative map $\mathcal{M}_{D,1}$ has a radius-dependent multifractal attractor. These multifractals are significantly more complicated (and beautiful!) than ones present in the literature, and we believe that it can be used to understand the properties of nonequilibrium dynamical systems. 

\section{Acknowledgement}
The author thanks Prof. Baidurya Bhattacharya of Indian Institute of Technology Kharagpur for his insightful comments.

\bibliography{paper}
\end{document}